\documentclass{PoS}

\usepackage{cite}
\usepackage{subfig}

\newcommand{\kbar}{\bar{k}}
\newcommand{\thetat}{\tilde{\theta}}
\newcommand{\Tr}{\textrm{Tr}\,}

\title{Glueball masses in 2+1 dimensional SU(N) gauge theories with twisted
boundary conditions}

\ShortTitle{Glueball masses in 2+1 dimensional SU(N) gauge theories with twisted b.c.}

\author{Margarita Garc\'ia P\'erez\\
        Instituto de F\'isica Te\'orica UAM/CSIC, Universidad Aut\'onoma de Madrid,
        E-28049-Madrid, Spain\\
        E-mail: \email{margarita.garcia@uam.es}}

\author{Antonio Gonz\'alez-Arroyo\\
        Instituto de F\'isica Te\'orica UAM/CSIC and Departamento de F\'isica
        Te\'orica, C-15, Universidad Aut\'onoma de Madrid, E-28049-Madrid,
        Spain\\
        E-mail: \email{antonio.gonzalez-arroyo@uam.es}}

\author{\speaker{Mateusz Kore\'n}\\
        Instituto de F\'isica Te\'orica UAM/CSIC, Universidad Aut\'onoma de Madrid,
        E-28049-Madrid, Spain\\
        E-mail: \email{mateusz.koren@csic.es}}

\author{Masanori Okawa\\
        Graduate School of Science, Hiroshima University, Higashi-Hiroshima,
        Hiroshima 739-8526, Japan\\
        E-mail: \email{okawa@sci.hiroshima-u.ac.jp}}

\abstract{We analyze 2+1 dimensional Yang-Mills theory regularized on a lattice
with twisted boundary conditions in the spatial directions. In previous work it
was shown that the observables in the non-zero electric flux sectors obey the
so-called $x$-scaling, i.e.\ depend only on the dimensionless variable $x\propto
NL/b$ and the angle $\thetat$ given by the parameters of the twist ($L$ being
the length of the spatial torus and $b$ the inverse 't Hooft coupling). It is
conjectured that this scaling is obeyed by all physical quantities. In this work
we extend the previous analyses to the zero electric flux (glueball) sector. We
study the mass of the lightest scalar glueball in two theories with different
$N$ but matching $x$ and $\thetat$ in a wide range of couplings from the
perturbative small-volume regime to the non-perturbative one. We find that the
results are consistent with the $x$-scaling hypothesis.}

\FullConference{The 32nd International Symposium on Lattice Field Theory,\\
    23-28 June, 2014\\
    Columbia University New York, NY}

\begin{document}

\section{Introduction}

Large-$N$ gauge theories ($N$ being the degree of the gauge group) exhibit a
number of peculiar features related to the fact that only the leading
contribution in the so-called $N$-counting rules survives in the $N\to\infty$
limit (see e.g.\ Refs.~\cite{man98,lpa13} for review). One particularly
interesting feature of the large-$N$ limit is the emergence of the so-called
large-$N$ equivalences, linking theories with different parameters such as the
matter content, gauge groups or the spacetime volume on which the theories are
defined \cite{lov82,kuy04,asv03}.

In fact, one of the most renowned examples of such equivalences is the
Eguchi-Kawai reduction \cite{EK82}, also known as volume reduction or volume
independence, which (in the lattice language) relates two $SU(N)$ gauge
theories: one defined on infinite lattice and the other defined on a toroidal
lattice of arbitrarily small size (including one spacetime lattice site).

The Eguchi-Kawai reduction requires that the center symmetry in the small-volume
theory remains unbroken, which turns out to be a non-trivial condition. One of
the ways to fulfill it is to use twisted boundary conditions in the spacetime
torus \cite{TEK1,TEK2,TEK3,TEK4}. Other methods include partial reduction (which
requires keeping the physical volume of the reduced model large enough, $\gtrsim
1\textrm{fm}^4$) \cite{KNN03}, the addition of adjoint fermions to the model
\cite{kuy07,BCD13,kor13} and the related idea of trace-deformed reduction
\cite{uny08}.

Note that the large-$N$ equivalences are only strictly true in the $N\to\infty$
limit. In this work we follow a slightly different approach and analyze the
interplay between finite $N$ and $L$, where $L$ denotes the size (in lattice
units) of the spatial torus on which the theory is defined.

In our particular case, we analyze 2+1-dimensional lattice gauge theory
(extensions to 3+1 dimensions are also possible, see Ref.~\cite{ggo14}) defined
on a spatial torus with twisted boundary conditions. The action of the model is
given by:
\begin{equation}
S = Nb
\sum_{n}\sum_{\mu\neq\nu}\left(N-z^*_{\mu\nu}(n)P_{\mu\nu}(n)\right),
\end{equation}
where $n$ runs over the lattice of size $L\times L\times T$, $b$ is the inverse
't Hooft coupling, $P_{\mu\nu}(n)$ is the plaquette and $z_{\mu\nu}$ is the
twist tensor equal to 1 except at corner plaquettes in (1,2)-plane where it is
equal to:
\begin{equation}
z_{ij}(n)=\exp\Big(i\epsilon_{ij}\frac{2\pi k}N\Big),
\end{equation}
where $k$ is an integer known as the magnetic flux. We also define integer
$\kbar$ as the modular multiplicative inverse of $k$:
\begin{equation}
k\kbar=1 \ (\textrm{mod }N).
\end{equation}
  
Our interest is motivated by the work of Ref.~\cite{GGO13} where this setup was
analyzed in the non-zero electric flux sector, which in the large volume
corresponds to the $k$-string tensions. In this work it was shown, both in
perturbation theory to all orders and in non-perturbative lattice calculations,
that the $k$-string tensions depend only on the dimensionless scaling
variable\footnote{Note that the perturbative calculations are done in the
continuum, the $x$ given here is equal to the continuum $x$ up to finite $a$
corrections.}:
\begin{equation}
x=\frac{NL}{4\pi b}
\end{equation}
and the angle determined by the coefficient $\kbar$:
\begin{equation}
\thetat = \frac{2\pi\bar{k}}{N}.
\end{equation}

Thus, for given twist\footnote{One has to scale $k$ and $\kbar$ with $N$
accordingly, much as in the TEK model \cite{TEK3,TEK4}, to avoid tachyonic
behaviour of the theory, see Ref.~\cite{GGO13}.}, the physics of the non-zero
flux sector depends on $N$ and $L$ only via the product $NL$. This can be
thought as a generalized volume reduction where volume and gauge group size can
be interchanged also at finite $N$.

The long-term goal of this work is to verify whether this fact holds also in the
zero electric flux sector (corresponding to the glueball and torelon spectrum).
In this paper we restrict ourselves to comparing the mass of the lightest scalar
glueball in two theories:
\begin{enumerate}
  \item {$N=5$, $L=14$, $\kbar=2$, corresponding to $NL=70$,
  $\thetat\approx2.513$}
  \item {$N=17$, $L=4$, $\kbar=7$, corresponding to $NL=68$.
  $\thetat\approx2.587$}
\end{enumerate}
The theories are chosen so that while the values of $N$ are vastly different,
the values of the parameters $x$ and $\thetat$, determining the physical
behaviour (at least in the non-zero electric flux sector), are close within a
couple percent. Thus, if the hypothesis that this behaviour extends to the glueball
sector is true, the values of the glueball mass should be approximately equal in
the two theories for all values of the coupling $b$ (corresponding to different
values of the scaling parameter $x$).

\section{Calculation}

The calculation of the glueball masses is performed in a wide range of
couplings\footnote{For sake of unified terminology, in this work we use the name
``glueball'' for the gluonic states with the quantum numbers of the glueball --
also in the weak-coupling, small-volume region where the states correspond to
(pairs of) non-contractible flux tubes and one might argue that the name
``torelon'' should be used.}. In particular we need to deal with several regions
in which the behaviour is widely different -- knowledge from earlier works
\cite{GGO13,TEP98} as well as some hindsight allow us to distinguish three such
regions:
\begin{enumerate}
  \item {Perturbative, small-volume region: $x\lesssim0.5$.}
  \item {Intermediate region: $0.5\lesssim x\lesssim3$.}
  \item {Large-volume region: $x\gtrsim3$. Note: to have better access to the
  large-volume region we also use the theories with the lattice torus size $L$
  doubled.}
\end{enumerate}

The extraction of the ground state for the glueballs is not an easy task and
requires variational analysis, using the solution of the Generalized Eigenvalue
Problem (called GEVP in the following).

The observables $O_i(t)$ we use are rectangular Wilson loops and squared moduli
of multi-winding spatial Polyakov loops $|\Tr P^{\,n}|^2$, projected to zero
momentum and angular momentum.

We employ three different levels of APE \cite{APE87} smearing on the loops (7,
14 and 21 steps, with smearing parameter $\alpha=0.475$) and, instead of
blocking, use Wilson loops of large sizes, trying, for given values of the
coupling, to choose the loop sizes whose correlators give good signal at large
time separations  (i.e.\ trying to follow the physical size of the glueball).
That includes, in particular using Wilson loops larger than the spatial extent
of the lattice for small $x$.

From the observables we construct the correlation matrix:
\begin{equation}
C_{ij}(t) = \sum_{t'}\langle O_i(t'+t)O_j(t') \rangle -
\langle O_i(t'+t)\rangle \langle O_j(t') \rangle,
\end{equation}
on which we solve the GEVP for $t_0=a$, $t_1=t_0+a$ (we find that in practice,
increasing the value of $t_0$ does not change the results significantly):
\begin{equation}
C(t_1)v = C(t_0)\lambda v.
\end{equation}
We then use the obtained eigenvectors $v$ to change the basis of $C(t)\to
\tilde{C}(t)$ for all values of $t$ and use the diagonal values of
$\tilde{C}(t)$ to extract the plateau ranges and subsequently perform fits on
the selected ranges.

One subtlety when fitting outside the large-volume region is that the
finite-temperature corrections turn out to play a sizeable role in the
correlators for the temporal extents $T$ of the lattices used ($T=72$, except
the large-volume region, where $T=36$). In perturbation theory one can show that
the correction is proportional to $\exp(-mT/2)$ where $m$ is the mass of the
lightest glueball, rather than $\exp(-mT)$; the factor 2 comes from the fact
that there is a single gluon propagating around the temporal torus which has
energy $m/2$. This forces us to include the effect of the constant term on the
fits. We do that using the midpoint-subtracted correlators \cite{ume07} in the
effective mass plots and the fits.

The correlation matrix typically consists of approximately $15$ operators. We
verify whether the basis allows for reliable GEVP solution by first solving it on
non-symmetrized correlation matrix -- the breakdown of the non-symmetrized GEVP
signals insufficient signal-to-noise ratio. The GEVP and basis change are done
in quadruple (128 bit) floating-point precision to avoid adding round-off errors
to the problem.

\section{Results}

The physics of the problem is significantly different in the three regions of
interest. In the large-volume (large-$x$) region the result is expected to be
close to the non-twisted large-volume calculations. In this region the operators
with largest overlap to the ground state are the contractible Wilson loops. The
choice of boundary conditions should be irrelevant in large volume thus we
expect that the results are consistent with \cite{TEP98}.

On the other hand, in the small-$x$ region, the expectations can be made using
perturbation theory. In the twisted theory, the leading contribution for the
lowest energy is \cite{GGO13}:
\begin{equation}
{b \cal{E} } =  {1 \over  x} - 2G(\thetat/2\pi) - {3 \over 4
\pi^2} \sin^2 (\thetat/2) \, ,
\end{equation}
where we have introduced the function:
\begin{equation}
G (z) = -{1 \over 16 \pi^2} \int_0^\infty
{dt \over \sqrt{ t}} \,
\Big(\theta_3(0,it) \big( \theta_3(0,it) -\theta_3(z,it)\big)- {1 \over
t}\Big).
\end{equation}
with $\theta_3$ the Jacobi Theta function. Also, the operator with the highest
overlap to the ground state is $|\Tr P^{\,\kbar}|^2$, corresponding to the pair
of (mutually conjugate, i.e.\ carrying momenta with opposite signs) gluonic
operators, each having the lowest possible momentum given by
\begin{equation}
|\vec{p}| = \frac{2\pi}{NL}\,.
\end{equation}

The hardest to analyze is the intermediate-$x$ region where there are no
theoretical expectations and where different states contribute to the result.
We expect level crossing in this region and many states with similar energies
may be found. We find that in this region it is necessary to include $|\Tr
P^{\,\kbar}|^2$, $|\Tr P|^2$ and the Wilson loops whose contribution to the
ground state changes as $x$ goes from the perturbative to the large-volume
regime.

\begin{figure}[tbp!]
\centering
\captionsetup[subfigure]{oneside,margin={1cm,0cm}}
\subfloat[]{\label{fig:mb} \includegraphics[width=12cm]{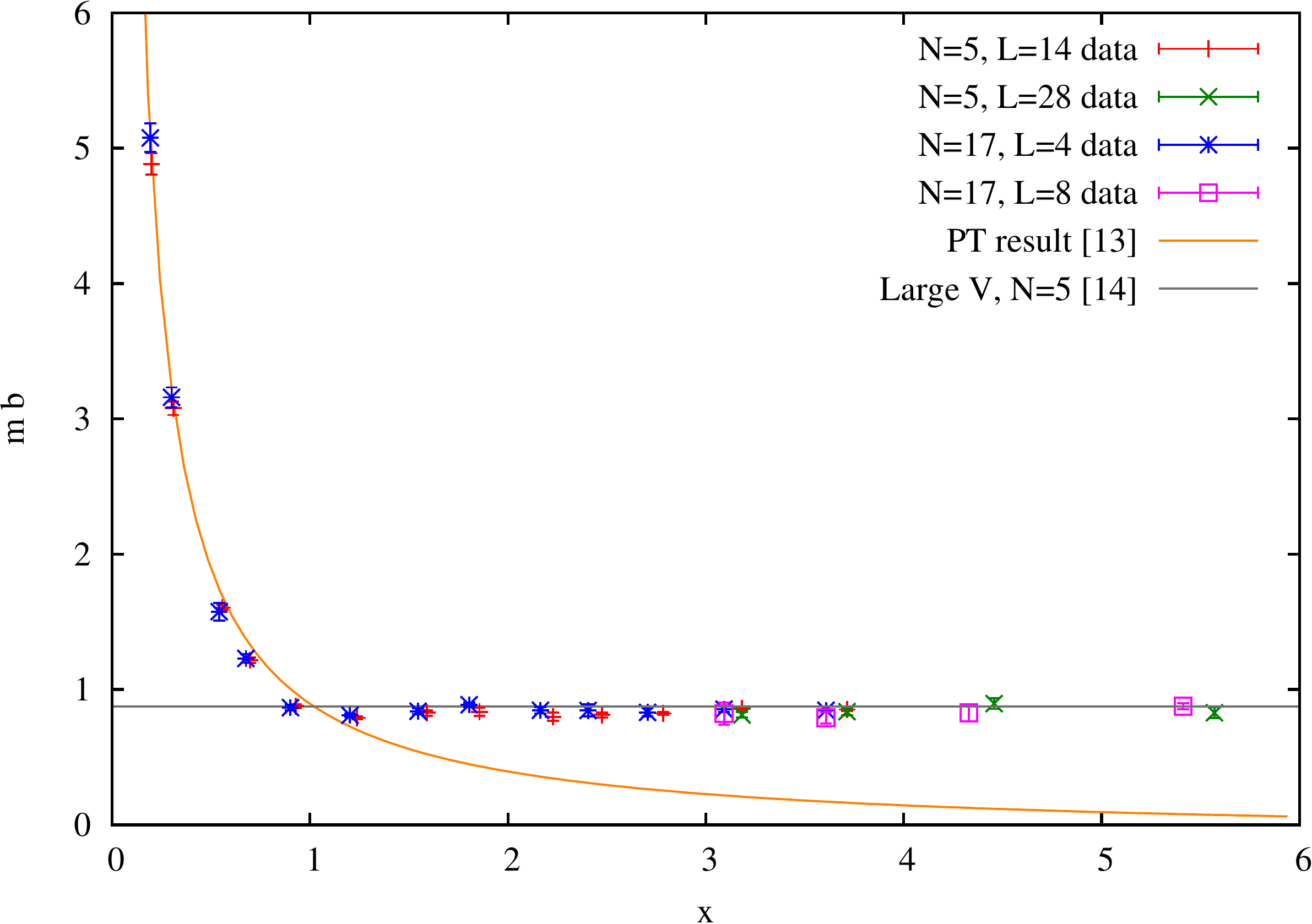}}\\
\captionsetup[subfigure]{oneside,margin={1.25cm,0cm}}
\subfloat[]{\label{fig:mln} \includegraphics[width=12.2cm]{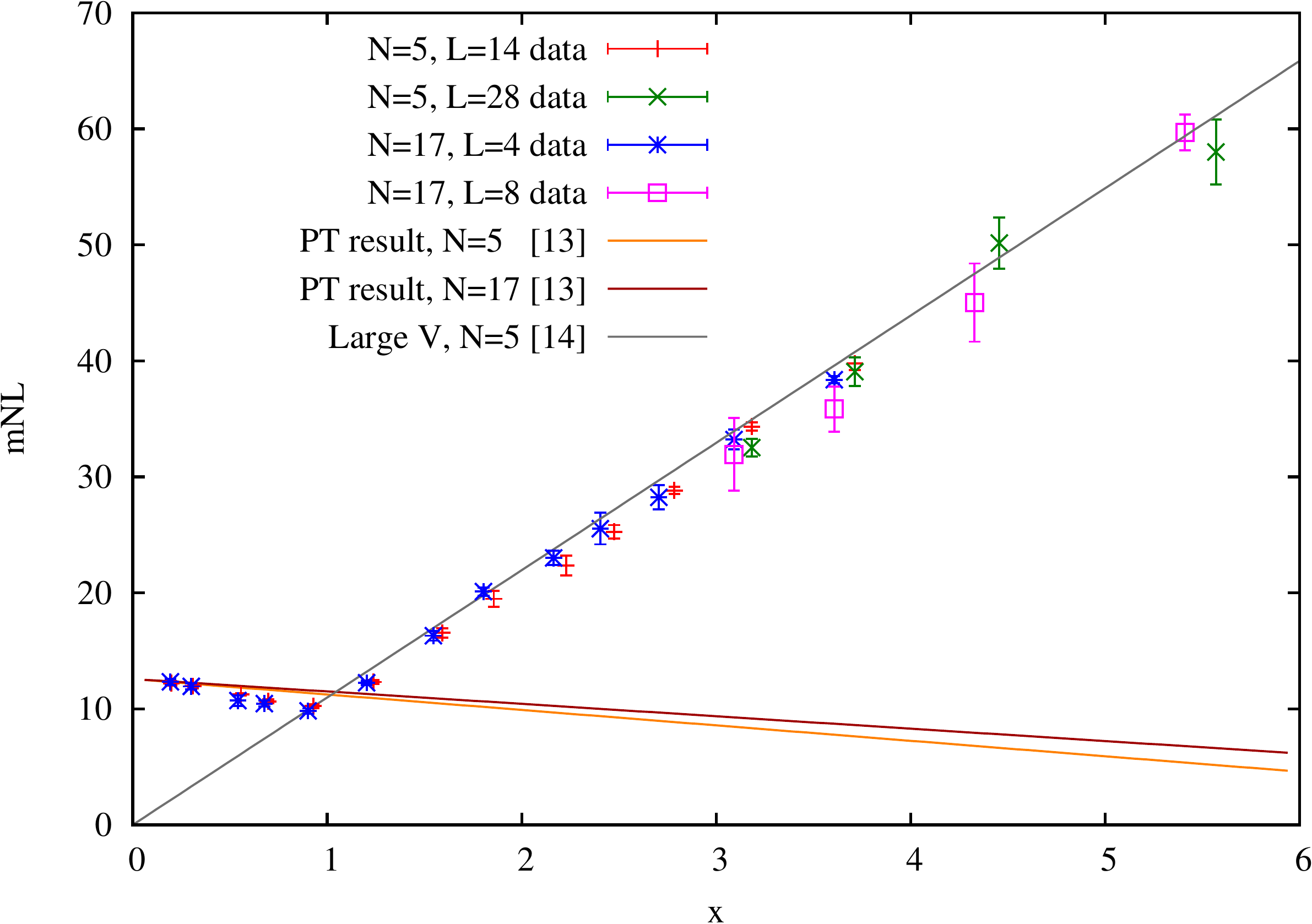}}

\caption{The results for the lightest scalar glueball mass as a function of the
$x$ variable. Two quantities are shown: Fig.~(a) shows the lattice results
rescaled times the inverse 't Hooft coupling and in Fig.~(b) the lattice results
are rescaled by a factor $NL$. The error bars are purely statistical, coming
from the jackknife estimation of errors from the correlated fits. In Fig.~(a)
only one perturbative result is drawn as the results for the two theories are
indistinguishable.}
\label{fig:results}
\end{figure}

The results, together with theoretical expectations, are presented in
Fig.~\ref{fig:results}. Fig.~\ref{fig:mb} presents the results in the units of
$mb=m/(g^2N)$ ($g^2$ has dimension of mass in 2+1 dimensions). On
Fig.~\ref{fig:mln} the same results are presented in different units, introduced
as following. The standard rescaling in finite volume analyses is done using the
control variable $z=mL$ \cite{lum84}. On the other hand, if the $x$-scaling
hypothesis is correct, we expect the data points to be governed rather by the
combination $mNL$, which is presented in Fig.~\ref{fig:mln}.

The plots show a rather striking agreement between the two theories. The results
for $N=5$ and $N=17$ are consistent within errors for all values of $x$
analyzed. This is a strong confirmation of the $x$-scaling hypothesis also in
the zero electric flux sector.

Another observation is that the change of behaviour between the perturbative and
large-volume like behaviour is rather abrupt. Even in the region $1\lesssim
x\lesssim3$ where the inclusion of the moduli of Polyakov loops in the GEVP
basis is necessary to get satisfying plateaux, the results follow the
large-volume value closely.

\section{Conclusions \& outlook}

We verified that the mass of the lightest scalar glueball in 2+1 dimensions with
twisted spatial boundary conditions is consistent between theories with $N=5$
and $N=17$ with matching values of the parameters $ x \propto NL/b$ and $\thetat
\propto \kbar /N$ for a very wide range of couplings, including all regions of
physical interest.

The equality of masses of the glueball states, belonging to the zero electric
flux sector, gives a strong support to the $x$-scaling hypothesis, which states
that the physics of twisted theories in 2+1 dimensions can be accurately
described by using solely two dimensionless parameters $x$ and $\thetat$, as was
already verified in the non-zero electric flux sector (corresponding to the
$k$-string tensions) in Ref.~\cite{GGO13}.

The analysis presented here can be improved in many ways. To reduce the
systematic errors the introduction of well-defined criteria for the choice of
plateau ranges is necessary. Choosing the plateaux ``by the eye'' is
particularly problematic in the presence of the midpoint subtraction procedure.
We plan to eliminate this difficulty by choosing the temporal extent $T$ to be
large enough to suppress the finite-temperature related constant term. This will
also allow to use the GEVP in the way done by ALPHA Collaboration which puts the
contamination by excited states under better control \cite{alp09}.

For the forthcoming publication we also plan to include another value of $N$ and
study the $\kbar$ dependence, as well as calculate the mass of the tensor
glueball.

\section*{Acknowledgments}
We acknowledge financial support from the MCINN grants FPA2012-31686 and
FPA2012-31880, the Comunidad Aut\'onoma de Madrid under the program HEPHACOS
S2009/ESP-1473, and the Spanish MINECO's ``Centro de Excelencia Severo Ochoa''
Programme under grant SEV-2012-0249. M. O. is supported by the Japanese MEXT
grant No 26400249. The numerical simulations were done on the HPC-clusters at
IFT.

\end{document}